\count100=1 
\def\doit#1#2{\ifcase#1\or#2\fi}

 2
\font\LARGE=cmr10 scaled \magstep 3
 
\font\cmsytenscaled=cmsy10 scaled \magstep 3

\def\Tab{(T^I)^{A B}} 
\def\doubletilde#1{{}{\buildrel{\mkern1mu_\approx\mkern-1mu}%
\over{#1}}{}}

\catcode`@=11 \catcode`@=12

\doit1{ 
\font\eightrm=cmr8 
\font\eightbf=cmbx8

\font\nineit=cmti9
\font\eightsl=cmss8 
\font\eightmus=cmmi8 
\def\smalltype{\let\rm=\eightrm \let\bf=\eightbf \let\it=\nineit
\let\sl=\eightsl \let\mus\eightmus 
\baselineskip=1.5pt minus .75pt\rm} 
\def\it{\fam\itfam\tenit} 
} 

\def\footnotes#1#2{\footnote{$^{\star#1)}$}{\smalltype #2}} 
\def\newpage{\vfill\eject}

\def\today{\ifcase\month\or January\or February\or March\or
April\or May\or June\or July\or August\or September\or 
October\or November\or December\fi \space\number\day, 
\number\year} 

\let\du=\d 

\def\a{\alpha} \def\b{\beta}  \def\d{\delta}
\def\e{\epsilon}  \def\g{\gamma}
   
\def\l{\lambda} \def\m{\mu} \def\n{\nu} \def\o{\omega}
  \def\r{\rho} \def\s{\sigma}
\def\t{\tau}   
   
\def\L{\Lambda}

\def\pmb#1{\setbox0=\hbox{${#1}$}%
   \kern-.025em\copy0\kern-\wd0
   \kern-.035em\copy0\kern-\wd0
   \kern.05em\copy0\kern-\wd0
   \kern-.035em\copy0\kern-\wd0
   \kern-.025em\box0 }


\def\bo{{\raise-.46ex\hbox{\large$\Box$}}} 

\def\pr{\prod} 

\def\TH{{\raise.2ex\hbox{$\displaystyle \bigodot$}\mskip-4.7mu %
\llap H \;}}
\def\face{{\raise.2ex\hbox{$\displaystyle \bigodot$}\mskip-2.2mu %
\llap {$\ddot
    \smile$}}} 

\def\sp#1{{}^{#1}} 
\def\sl#1{\rlap{\hbox{$\mskip 1 mu /$}}#1} 

 %
 %

\def\Tilde#1{{\widetilde{#1}}\hskip 0.015in} 
\def\Hat#1{\widehat{#1}} 
\def\Bar#1{\overline{#1}} 
\def\leftrightarrowfill{$\mathsurround=0pt \mathord\leftarrow
 \mkern-6mu
    \cleaders\hbox{$\mkern-2mu \mathord- \mkern-2mu$}\hfill
    \mkern-6mu \mathord\rightarrow$}
\def\dvec#1{\vbox{\ialign{##\crcr
    \leftrightarrowfill\crcr\noalign{\kern-1pt\nointerlineskip}
    $\hfil\displaystyle{#1}\hfil$\crcr}}} 

\def\frac#1#2{{\textstyle{#1\over\vphantom2\smash{\raise.20ex
    \hbox{$\scriptstyle{#2}$}}}}} 
\def\sfrac#1#2{{\vphantom1\smash{\lower.5ex\hbox{\small$#1$}}%
\over\vphantom1\smash{\raise.4ex\hbox{\small$#2$}}}}
\def\bfrac#1#2{{\vphantom1\smash{\lower.5ex\hbox{$#1$}}\over
    \vphantom1\smash{\raise.3ex\hbox{$#2$}}}} 
\def\afrac#1#2{{\vphantom1\smash{\lower.5ex\hbox{$#1$}}\over#2}} 


\newskip\humongous \humongous=0pt plus 1000pt minus 1000pt
\def\caja{\mathsurround=0pt}

\newif\ifdtup
\def\panorama{\global\dtuptrue \openup2\jot \caja
    \everycr{\noalign{\ifdtup \global\dtupfalse
    \vskip-\lineskiplimit \vskip\normallineskiplimit
    \else \penalty\interdisplaylinepenalty \fi}}}
\def\li#1{\panorama \tabskip=\humongous 
    \halign to\displaywidth{\hfil$\displaystyle{##}$
    \tabskip=0pt&$\displaystyle{{}##}$\hfil
    \tabskip=\humongous&\llap{$##$}\tabskip=0pt
    \crcr#1\crcr}}

\doit0{
\def\ref#1{$\sp{#1)}$}
}

\hsize=6in 
\parskip=\medskipamount 
\lineskip=0pt 
\abovedisplayskip=1em plus.3em minus.5em 
\belowdisplayskip=1em plus.3em minus.5em 
\abovedisplayshortskip=.5em plus.2em minus.4em 
\belowdisplayshortskip=.5em plus.2em minus.4em 

\def\endtitle{\end{quotation}\newpage} 

\def\sect#1{\bigskip\medskip \goodbreak \noindent{\bf {#1}} %
\nobreak \medskip}
\def\refs{\sect{References} \footnotesize \frenchspacing \parskip=0pt}
\def\hang{\hangindent\parindent}
\def\textindent#1{\indent\llap{#1\enspace}\ignorespaces} 
\def\Item{\par\hang\textindent}

\def\[{\lfloor{\hskip 0.35pt}\!\!\!\lceil}
\def\]{\rfloor{\hskip 0.35pt}\!\!\!\rceil}

\def\Lag{{\cal L}} 
\def\du#1#2{_{#1}{}^{#2}}

\def\calR{{\cal R}}

\def\rma{{\rm a}} \def\rmb{{\rm b}}

\def\plpl{{+\!\!\!\!\!{\hskip 0.009in}%
{\raise-1.0pt\hbox{$_+$}} {\hskip 0.0008in}}}
\def\mimi{{-\!\!\!\!\!{\hskip 0.009in}%
{\raise-1.0pt\hbox{$_-$}} {\hskip 0.0008in}}}

\def\pl#1#2#3{Phys.~Lett.~{\bf {#1}B} (19{#2}) #3}
\def\np#1#2#3{Nucl.~Phys.~{\bf B{#1}} (19{#2}) #3}
\def\prl#1#2#3{Phys.~Rev.~Lett.~{\bf #1} (19{#2}) #3}
\def\pr#1#2#3{Phys.~Rev.~{\bf D{#1}} (19{#2}) #3}
\def\cqg#1#2#3{Class.~and Quant.~Gr.~{\bf {#1}} (19{#2}) #3}
\def\cmp#1#2#3{Comm.~Math.~Phys.~{\bf {#1}} (19{#2}) #3}
\def\jmp#1#2#3{Jour.~Math.~Phys.~{\bf {#1}} (19{#2}) #3}
\def\ap#1#2#3{Ann.~of Phys.~{\bf {#1}} (19{#2}) #3}
\def\prep#1#2#3{Phys.~Rep.~{\bf {#1}C} (19{#2}) #3}

\def\ijmp#1#2#3{Int.~Jour.~Mod.~Phys.~{\bf A{#1}} (19{#2}) #3}

\def\ibid#1#2#3{{\it ibid.}~{\bf {#1}} (19{#2}) #3}

\def\mpl#1#2#3{Mod.~Phys.~Lett.~{\bf A{#1}} (19{#2}) #3}

\font\texttt=cmr10 
\def\hepth#1{{\texttt hep-th/{#1}}}

\def\prn#1#2#3{Phys.~Rev.~{\bf D{#1}} (20{#2}) #3}

\def\epj#1#2#3{Eur.~Phys.~Jour.~{\bf C{#1}} (20{#2}) {#3}}

\def\<<{<\!\!<} \def\>>{>\!\!>}
\def\Check#1{{\raise-1.0pt\hbox{\LARGE\v{}}{\hskip -10pt}{#1}}}

\def\eqques{{~\,={\hskip -11.5pt}\raise -1.8pt\hbox{\large ?}
{\hskip 4.5pt}}{}}

\def\fracmm#1#2{\,{{#1}\over{#2}}\,}

\def\frac#1#2{{\textstyle{#1\over\vphantom2\smash{\raise -.20ex
    \hbox{$\scriptstyle{#2}$}}}}} 

\def\scst{\scriptstyle}

\def\.{.$\,$}
\def\-{{\hskip 1.5pt}\hbox{-}}

\def\low#1{\hskip0.01in{\raise -2.3pt\hbox{${\hskip 1.0pt}\!_{#1}$}}}
\def\Low#1{\hskip0.015in{\raise -3pt\hbox{$\!_{#1}$}}}
\def\lowlow#1{\hskip0.01in{\raise -7pt%
\hbox{${\hskip1.0pt} \hskip-2pt_{#1}$}}}

\font\tenmib=cmmib10
\font\sevenmib=cmmib10 at 7pt 
\font\fivemib=cmmib10 at 5pt 
\font\tenbsy=cmbsy10
\font\sevenbsy=cmbsy10 at 7pt 
\font\fivebsy=cmbsy10 at 5pt 
\def\BMfont{\textfont0\tenbf \scriptfont0\sevenbf
                  \scriptscriptfont0\fivebf
        \textfont1\tenmib \scriptfont1\sevenmib
                   \scriptscriptfont1\fivemib
        \textfont2\tenbsy \scriptfont2\sevenbsy
                   \scriptscriptfont2\fivebsy}
\def\rlx{\relax\leavevmode}
\def\BM#1{\rlx\ifmmode\mathchoice
              {\hbox{$\BMfont#1$}}
              {\hbox{$\BMfont#1$}}
              {\hbox{$\scriptstyle\BMfont#1$}}
              {\hbox{$\scriptscriptstyle\BMfont#1$}}
         \else{$\BMfont#1$}\fi}

\font\tenmib=cmmib10
\font\sevenmib=cmmib10 at 7pt 
\font\fivemib=cmmib10 at 5pt 
\font\tenbsy=cmbsy10
\font\sevenbsy=cmbsy10 at 7pt 
\font\fivebsy=cmbsy10 at 5pt 
\def\BMfont{\textfont0\tenbf \scriptfont0\sevenbf
                  \scriptscriptfont0\fivebf
        \textfont1\tenmib \scriptfont1\sevenmib
                   \scriptscriptfont1\fivemib
        \textfont2\tenbsy \scriptfont2\sevenbsy
                   \scriptscriptfont2\fivebsy}
\def\BM#1{\rlx\ifmmode\mathchoice
              {\hbox{$\BMfont#1$}}
              {\hbox{$\BMfont#1$}}
              {\hbox{$\scriptstyle\BMfont#1$}}
              {\hbox{$\scriptscriptstyle\BMfont#1$}}
         \else{$\BMfont#1$}\fi}

\def\inbar{\vrule height1.5ex width.4pt depth0pt}
\def\sinbar{\vrule height1ex width.35pt depth0pt}
\def\ssinbar{\vrule height.7ex width.3pt depth0pt}

\def\ZZ{{}Z {\hskip -6.0pt} Z{}}
\def\Ik{\rlx{\rm I\kern-.18em k}} 
\def\IC{\rlx\leavevmode
         \ifmmode\mathchoice
            {\hbox{\kern.33em\inbar\kern-.3em{\rm C}}}
            {\hbox{\kern.33em\inbar\kern-.3em{\rm C}}}
            {\hbox{\kern.28em\sinbar\kern-.25em{\rm C}}}
            {\hbox{\kern.25em\ssinbar\kern-.22em{\rm C}}}
         \else{\hbox{\kern.3em\inbar\kern-.3em{\rm C}}}\fi}
\def\IP{\rlx{\rm I\kern-.18em P}}
\def\IR{\rlx{\rm I\kern-.18em R}}
\def\IN{\rlx{\rm I\kern-.20em N}}
\def\Ione{\rlx{\rm 1\kern-2.7pt l}}

%

\newbox\leftpage \newdimen\fullhsize \newdimen\hstitle%
\newdimen\hsbody%
\tolerance=1000\hfuzz=2pt
\catcode`\@=11 
\hsbody=\hsize \hstitle=\hsize 

\def\nolabels{\def\wrlabeL##1{}\def\eqlabeL##1{}%
\def\reflabeL##1{}}
\def\writelabels{\def\wrlabeL##1{\leavevmode%
\vadjust{\rlap{\smash%
{\line{{\escapechar=` \hfill\rlap{\sevenrm\hskip.03in\string##1}}}}}}}%
\def\eqlabeL##1{{\escapechar-1%
\rlap{\sevenrm\hskip.05in\string##1}}}%
\def\reflabeL##1{\noexpand\llap{\noexpand%
\sevenrm\string\string%
\string##1}}} \nolabels
%
\global\newcount\secno \global\secno=0 \global\newcount\meqno
\global\meqno=1
\def\newsec#1{\global\advance\secno by1\message{(\the\secno. #1)}
\global\subsecno=0\eqnres@t\noindent{\bf\the\secno. #1}
\writetoca{{\secsym} {#1}}\par\nobreak\medskip\nobreak}
\def\eqnres@t{\xdef\secsym{\the\secno.}\global\meqno=1
\bigbreak\bigskip}
\def\sequentialequations{\def\eqnres@t{\bigbreak}}\xdef\secsym{}
\global\newcount\subsecno \global\subsecno=0
\def\subsec#1{\global\advance\subsecno by1%
\message{(\secsym\the\subsecno.%
 #1)}
\ifnum\lastpenalty>9000\else\bigbreak\fi
\noindent{\it\secsym\the\subsecno. #1}\writetoca{\string\quad
{\secsym\the\subsecno.} {#1}}\par\nobreak\medskip\nobreak}
\def\appendix#1#2{\global\meqno=1\global\subsecno=0%
\xdef\secsym{\hbox{#1.}} \bigbreak\bigskip\noindent{\bf Appendix
#1. #2}\message{(#1. #2)} \writetoca{Appendix {#1.}
{#2}}\par\nobreak\medskip\nobreak}
\def\eqnn#1{\xdef #1{(\secsym\the\meqno)}%
\writedef{#1\leftbracket#1}%
\global\advance\meqno by1\wrlabeL#1}
\def\eqna#1{\xdef #1##1{\hbox{$(\secsym\the\meqno##1)$}}
\writedef{#1\numbersign1\leftbracket#1{\numbersign1}}%
\global\advance\meqno by1\wrlabeL{#1$\{\}$}}
\def\eqn#1#2{\xdef #1{(\secsym\the\meqno)}%
\writedef{#1\leftbracket#1}%
\global\advance\meqno by1$$#2\eqno#1\eqlabeL#1$$}
%
\newskip\footskip\footskip8pt plus 1pt minus 1pt
\font\smallcmr=cmr5
\def\footnotefont{\smallcmr}
\def\f@t#1{\footnotefont #1\@foot}
\def\f@@t{\baselineskip\footskip\bgroup\footnotefont\aftergroup%
\@foot\let\next}
\setbox\strutbox=\hbox{\vrule height9.5pt depth4.5pt width0pt} %
\global\newcount\ftno \global\ftno=0
\def\foot{\global\advance\ftno by1\footnote{$^{\the\ftno}$}}
%
\newwrite\ftfile
\def\footend{\def\foot{\global\advance\ftno by1\chardef\wfile=\ftfile
$^{\the\ftno}$\ifnum\ftno=1\immediate\openout\ftfile=foots.tmp\fi%
\immediate\write\ftfile{\noexpand\smallskip%
\noexpand\item{f\the\ftno:\ }\pctsign}\findarg}%
\def\footatend{\vfill\eject\immediate\closeout\ftfile{\parindent=20pt
\centerline{\bf Footnotes}\nobreak\bigskip\input foots.tmp }}}
\def\footatend{}
\global\newcount\refno \global\refno=1
\newwrite\rfile
\def\ref{[\the\refno]\nref}%
\def\nref#1{\xdef#1{[\the\refno]}\writedef{#1\leftbracket#1}%
\ifnum\refno=1\immediate\openout\rfile=refs.tmp\fi%
\global\advance\refno by1\chardef\wfile=\rfile\immediate%
\write\rfile{\noexpand
\Item{#1}\reflabeL{#1\hskip-10pt}\pctsign}%
\findarg\hskip10.0pt}%
\def\findarg#1#{\begingroup\obeylines\newlinechar=`\^^M\pass@rg}
{\obeylines\gdef\pass@rg#1{\writ@line\relax #1^^M\hbox{}^^M}%
\gdef\writ@line#1^^M{\expandafter\toks0%
\expandafter{\striprel@x #1}%
\edef\next{\the\toks0}\ifx\next\em@rk\let\next=\endgroup%
\else\ifx\next\empty%
\else\immediate\write\wfile{\the\toks0}%
\fi\let\next=\writ@line\fi\next\relax}}
\def\striprel@x#1{} \def\em@rk{\hbox{}}
\def\lref{\begingroup\obeylines\lr@f}
\def\lr@f#1#2{\gdef#1{\ref#1{#2}}\endgroup\unskip}

\def\addref#1{\immediate\write\rfile{\noexpand\item{}#1}} 
%

%


%
\def\startrefs#1{\immediate\openout\rfile=refs.tmp\refno=#1}
\def\xref{\expandafter\xr@f}\def\xr@f[#1]{#1}
\def\refs#1{\count255=1[\r@fs #1{\hbox{}}]}
\def\r@fs#1{\ifx\und@fined#1\message{reflabel %
\string#1 is undefined.}%
\nref#1{need to supply reference \string#1.}\fi%
\vphantom{\hphantom{#1}}\edef\next{#1}%
\ifx\next\em@rk\def\next{}%
\else\ifx\next#1\ifodd\count255\relax\xref#1\count255=0\fi%
\else#1\count255=1\fi\let\next=\r@fs\fi\next}

\newwrite\ffile\global\newcount\figno \global\figno=1
\doit0{
\def\fig{fig.~\the\figno\nfig}
\def\nfig#1{\xdef#1{fig.~\the\figno}%
\writedef{#1\leftbracket fig.\noexpand~\the\figno}%
\ifnum\figno=1\immediate\openout\ffile=figs.tmp%
\fi\chardef\wfile=\ffile%
\immediate\write\ffile{\noexpand\medskip\noexpand%
\item{Fig.\ \the\figno. }
\reflabeL{#1\hskip.55in}\pctsign}\global\advance\figno
by1\findarg}
\def\xfig{\expandafter\xf@g}\def\xf@g fig.\penalty\@M\ {}
\def\figs#1{figs.~\f@gs #1{\hbox{}}}
\def\f@gs#1{\edef\next{#1}\ifx\next\em@rk\def\next{}\else
\ifx\next#1\xfig #1\else#1\fi\let\next=\f@gs\fi\next}
}

\newwrite\lfile
{\escapechar-1\xdef\pctsign{\string\%}\xdef\leftbracket{\string\{}
\xdef\rightbracket{\string\}}\xdef\numbersign{\string\#}}

\def\writestop{\def\writestoppt%
{\immediate\write\lfile{\string\pageno%
\the\pageno\string\startrefs\leftbracket\the\refno\rightbracket%
\string\def\string\secsym\leftbracket\secsym\rightbracket%
\string\secno\the\secno\string\meqno\the\meqno}%
\immediate\closeout\lfile}}
\def\writestoppt{}\def\writedef#1{}
\def\seclab#1{\xdef #1{\the\secno}\writedef{#1\leftbracket#1}%
\wrlabeL{#1=#1}}
\def\subseclab#1{\xdef #1{\secsym\the\subsecno}%
\writedef{#1\leftbracket#1}\wrlabeL{#1=#1}}
\newwrite\tfile \def\writetoca#1{}
\def\leaderfill{\leaders\hbox to 1em{\hss.\hss}\hfill}
\def\writetoc{\immediate\openout\tfile=toc.tmp
   \def\writetoca##1{{\edef\next{\write\tfile{\noindent ##1
   \string\leaderfill {\noexpand\number\pageno} \par}}\next}}}

%
\catcode`\@=12 
%

\countdef\pageno=0 \pageno=1
\newtoks\headline \headline={\hfil}
\newtoks\footline
\footline={\tenrm\folio} 
\def\folio{\ifnum\pageno<0 \romannumeral-\pageno
\else{\hss\number\pageno\hss}\fi}

\def\advancepageno{\ifnum\pageno<0 \global\advance\pageno by -1
\else\global\advance\pageno by 1 \fi}
\newif\ifraggedbottom



\def\circle#1{$\bigcirc{\hskip-9pt}\raise-1pt\hbox{#1}$}

\def\eqdot{~{\buildrel{\hbox{\LARGE .}} \over =}~}

\def\eqques{~{\buildrel ? \over =}~}

\def\binomial#1#2{\left(\,{\buildrel
{\raise4pt\hbox{$\displaystyle{#1}$}}\over
{\raise-6pt\hbox{$\displaystyle{#2}$}}}\,\right)}

\font\smallcmr=cmr6 scaled \magstep2 
scaled \magstep 1 \font\largetitle=cmr17 scaled \magstep1

\font\large=cmr10 scaled \magstep3

\def\alephnull{\aleph_0}



\def\neq{\not=}

\font\smallcmr=cmr6 scaled \magstep2

\def\fracmm#1#2{{{#1}\over{#2}}}

\topskip 0.2in
 
\baselineskip 20.0pt
\vsize=9.0in
\hsize=6.5in
\pagedepth=-5in

\tolerance=10000
\magnification=1200

\pageno=1

\doit0{
{\bf Preliminary Version (FOR YOUR EYES ONLY!) \hfill \today}
} 
\vskip -0.03in

\doit1{
\rightline{hep-th/0504097}  
\vskip -0.1in 
}
\rightline{CSULB--PA--05--1} 
\vskip -0.1in 

\vskip 0.2in
\vskip 0.35in

\centerline{\largetitle\hbox{\cmsytenscaled @}$_0\, $%
-$\,$\large Hypergravity in Three-Dimensions}

\bigskip

\baselineskip 9pt

\vskip 0.18in

\centerline{Hitoshi ~N{\smallcmr ISHINO}%
\footnotes1{E-Mail: hnishino@csulb.edu}
~and ~Subhash ~R{\smallcmr AJPOOT}%
\footnotes2{E-Mail: rajpoot@csulb.edu} 
}

\bigskip
\centerline{\it Department of Physics \& Astronomy}
\vskip 0.03in
\centerline{\it California State University} 
\vskip 0.03in
\centerline{\it 1250 Bellflower Boulevard} 
\vskip 0.03in
\centerline{\it Long Beach, CA 90840} 
\bigskip

\vskip 2.0 in

\centerline{\bf Abstract}
\bigskip

\vskip 0.1in

\baselineskip 14pt 

We construct hypergravity theory in three-dimensions with the
gravitino $~\psi\du{\m m_1\cdots m_n} A$~ with an   
arbitrary half-integral spin $~n+3/2$, carrying
also the index $~{\scst A}$~ for certain real representations of any
gauge group $~G$.  The possible real  representations are restricted by
the condition that the matrix representation of all the generators are
antisymmetric:  $~ (T^I)^{A B} = - (T^I)^{B A}$.  
Since such a real representation can be
arbitrarily  large,  this implies $~\alephnull\-$hypergravity  with
infinitely many ($\alephnull$) extended local hypersymmetries.

\vskip 0.35in 
\leftline{\smalltype  PACS:  ~04.65.+e, 11.30.Pb, 02.40.Pc, 11.15.-q} 
\vskip -0.03in 
\leftline{\smalltype  Key Words:  Supergravity, 
Hypergravity, Hypersymmetry, Topological Field Theory, 
Chern-Simons,} 
\vskip -0.05in 
\leftline{\hskip 0.65in\smalltype  BF-Theory, 
Three-Dimensions } 
\vfill\eject

\baselineskip 18.0pt


\pageno=2
\leftline{\bf 1.~~Introduction}  

It is well-known that a graviton in three-dimensions (3D) has zero
physical degree of freedom.  This is because  the conventional counting
for a symmetric  traceless tensor for transverse components in 3D gives
zero:  $~(3-2)(4-2)/2 - 1 = 0$.  Similarly, the gravitino has also no
physical degree of freedom:  $~(3-3)\times 2=0$.  Therefore,  the
multiplet of supergravity has ~$0+0$~ physical degrees of freedom.  

This fact leads to the interesting concept of `hypergravity' 
\ref\hypergravity{C.~Aragone and S.~Deser, \pl{86}{79}{161}.}  
in 3D with a gravitino with spin 5/2 or higher
\ref\ad{C.~Aragone and S.~Deser, \cqg{1}{84}{L9}.}.  
In 4D, on the other hand, it is difficult to formulate consistent
hypergravity,   due to the problem with the free indices of a gravitino 
field equation whose divergences do not vanish on non-trivial 
backgrounds (Velo-Zwanziger disease) 
\ref\vz{G.~Velo and D.~Zwanziger, \pr{186}{69}{1337}.}.  
In contrast, this situation is drastically improved in 3D due to the zero
physical degree of freedom of  the gravitino.  The first hypergravity
theory in 3D was given in \ad, where it was shown that consistent
hypergravity indeed exists with the gravitino $~\psi_{\m m_1\cdots
m_n}$~  carrying spin $~n+3/2$, where the indices $~{\scst
m_1\cdots m_n}$~ are totally symmetric $~\gamma\-$traceless 
Lorentz indices.  Interacting models of 
higher spin gauge fields with extended hypersymmetry 
in anti-De Sitter 3D have been also developed  
\ref\pv{S.F.~Prokushkin and M.A.~Vasiliev, \np{545}{99}{385}, 
hep-th/9806236.}.  

Independent of this, there has been a different development 
about 3D physics related to Chern-Simons theories  
\ref\djt{S.~Deser, R.~Jackiw and S.~Templeton, 
\prl{48}{82}{975}; \ap{140}{82}{372}, Erratum-\ibid{185}{88}{406}.}%
\ref\hagen{C.R.~Hagen, \ap{157}{84}{342}; \pr{31}{85}{331}.}%
\ref\witten{E.~Witten, \cmp{121}{89}{351};
K.~Koehler, F.~Mansouri, C.~Vaz and L.~Witten,
\mpl{5}{90}{935}; \jmp{32}{91}{239}.}%
\ref\carlip{S.~Carlip, J.~Korean Phys.~Soc.~{\bf 28} (1995)
S447, gr-qc/9503024, {\it and references therein};  
{\it `Quantum Gravity in 2+1 Dimensions'}, Cambridge
University Press (1998).}.  
It had been known for some time that 
arbitrarily many ($\alephnull$) supersymmetries 
\ref\at{A.~Achucarro and P.K.~Townsend,
\pl{180}{86}{89}; \pl{229}{89}{383}.}%
\ref\ngscs{H.~Nishino and S.J.~Gates, Jr., 
\ijmp{8}{93}{3371}.}%
\ref\ngaleph{H.~Nishino and S.J.~Gates, Jr.,  
hep-th/9606090, \np{480}{96}{573};
H.~Nishino and S.~Rajpoot, \prn{67}{03}{025009},  
hep-th/0209106.}%
\ref\nps{W.G.~Ney, O.~Piguet and W.~Spalenza, 
\epj{36}{04}{245}, hep-th/0312193.}  
can be accommodated in supersymmetric Chern-Simons 
theory.   Typical examples are $~OSp(p|2;\IR)\times OSp(q|2;\IR)$~ \at\ 
or $~SO(N)$~ \ngscs\ which can be arbitrarily large.  
In our recent paper 
\ref\nralephnull{H.~Nishino and S.~Rajpoot, \prn{70}{04}{027701}, 
\hepth{0402111}.},   
we have shown that the gravitino, coupling to 
locally supersymmetric Chern-Simons theory in 3D, can 
be in the adjoint representation of any gauge group $~G$, 
with the relationship $~N = \hbox{dim}\, G$, instead of 
limited groups, such as $~OSp$~ or $~SO\-$type gauge groups \at\ngscs.  

In this paper, we combine these two developments in 3D.  
Namely, we generalize massless hypergravity \ad\ further to a 
system, where the gravitino $~\psi\du{m m_1\cdots m_n}A$~ carries
the real representation index $~{\scst A}$~ of an arbitrary gauge group
$~G$, such that the matrix representations of all the generators are
antisymmetric.  Since such a representation can be 
arbitrarily large, this means that we have 
`$\alephnull\-$hypergravity' with infinitely many extended
hypersymmetries.   

In the next section, we will first give the action $~I_0$~  which is a
generalization  of our previous system \nralephnull\ to the new system
where the gravitino has the index structures $~\psi\du{\scst m_1 
m_2\cdots m_n}A$.  We 
next consider additional Chern-Simons actions ~$I_{R \o}$~ and $~I_{G
B}$, where $~I_{R \o}$~ is a Lorentz Chern-Simons term, while
$~I_{G B}$~ is a $~G B\-$type Chern-Simons term.  Fortunately, the
invariance of the total action $~I_{\rm tot} \equiv I_0 + I_{R \o} 
+ I_{G B}$~ under hypersymmetry is confirmed under a 
slight modification of the hypersymmetry transformation rule.

\bigskip\bigskip\medskip 

\leftline{\bf 2.~~Total Lagrangian and 
$~\alephnull\-$Hypersymmetries} 

Our field content of the system is $~(e\du\m m, \psi_{m_1\cdots m_n}{}
^A, A\du\m  I, B\du\m I, C\du\m I, \l_{m_1\cdots m_n}{}^A)$~ which is 
similar to 
\ref\ngalephnull{H.~Nishino and S.J.~Gates, \np{480}{96}{573}, 
\hepth{9606090}.}\nralephnull.  
Our most important ingredient here, however, is the representation of
the gravitino $~\psi$~ and the gaugino $~\l$.  The indices 
$~_{m_1\cdots m_n}$~ are totally symmetric Lorentz indices carrying
spin $~n+3/2$~ as in \ad, and the gravitino has the additional constraint of 
$~\g\-$tracelessness as in \ad:   
$$ \li{ & \g^{m_1} \psi\du{\m m_1 m_2\cdots m_n} A = 0 ~~.
&(2.1) \cr} $$ 
The superscript $~{\scst A}$~ is a `collective' index for any 
real representation of an arbitrary gauge group $~G$, 
satisfying the condition 
$$ \li{ & (T^I)^{A B} = - (T^I)^{B  A}~~~~~ ~~~
     ({\scst I~=~1,~2,~\cdots, ~\hbox{dim}\,G})~~   
&(2.2) \cr } $$ 
for the generators $~T^I$~ 
of $~G$.   Typical examples of representations satisfying (2.2) are 
\Item{(i)} Vectorial Representation of $~SO(M): ~{\scst A~\equiv~a}, ~~ 
     {\scst B~\equiv~b}, ~~ (T^I)^{A B} = (T^I)^{a b} 
          = - (T^I)^{b a}. $  
\Item{(ii)} Adjoint Representation of $~^\forall G:  
       ~{\scst A~\equiv~J}, ~~ 
     {\scst B~\equiv~K}, ~~ (T^I)^{A B} = - f^{I J K}. $ 
\Item{(iii)} Totally Symmetric $~m$~ Vectorial Indices 
of $~SO(M): ~ {\scst A ~\equiv~ (a_1\cdots a_m)}, ~~ 
    {\scst B ~\equiv~ (b_1\cdots b_m)},$ 
\vskip -0.41in 
$$ \li{&  ~~~\,  (T^I)^{a_1\cdots a_m, b_1\cdots b_m} 
     = + m (T^I)^{(a_1|}{}_{(b_1|} \d\du{|b_2|}{|a_2|} \cdots 
         \d\du{|b_m)}{|a_m)}
       = -(T^I)^{b_1\cdots b_m, a_1\cdots a_m} {~~.  ~~~~~ ~~~~~}  
&(2.3)  \cr } $$ 
Note that the indices $~{\scst A,~B}$~ are `collective' indices in the case
(iii).  This case is analogous to the totally symmetric Lorentz  indices
$~{\scst m_1\cdots m_n}$~ on the gravitino \ad.  

We propose an action $~I_0 \equiv \int d^3 x\, \Lag_0$~ with 
the lagrangian  
$$ \li{ \Lag_0 = \, & - \frac 14 e R(\o)
      + \frac 12 \e^{\m\n\r} \big(\Bar\psi\du{\m(n)} A D_\n (\o, B) 
      \psi\du\r{(n) A} \big) 
     + \frac 12 g \e^{\m\n\r} C\du\m I G_{\n\r}{}^I \cr 
& + \frac 12 g h \e^{\m\n\r} \big(F\du{\m\n} I A\du\r I 
     - \frac  13 g f^{I J K} A\du\m I A\du\n J A\du \r K \big) 
     + \frac 12 g h e \big(\Bar\l\du{(n)}A \l^{(n) A} \big) ~~, 
&(2.4) \cr } $$
where the suffix $~{\scst (n)}$~ stands for the totally symmetric 
$~n$~ Lorentz indices  $~{\scst m_1\cdots m_n}$~ in order to save 
space.  The $~g$~ and $~h$~ are real coupling constants, but are 
related to each other for certain gauge groups, 
as will be seen shortly in (2.23).  
The third $~C G\-$term is a kind of $~B F\-$terms used in 
topological field theory  
\ref\tft{{\it See, e.g.,} D.~Birmingham, M.~Blau, 
M.~Rakowski and G.~Thompson, \prep{209}{91}{129}.},  
while the last line is a hypersymmetric Chern-Simons term 
with the gaugino mass term \ngscs.  
In the case when the suffix $~{\scst A}$~ needs a non-trivial metric, 
we have to distinguish the superscript and subscript.\footnotes3{Even 
though we consider first compact groups for $~{\scst G}$, we can 
generalize it to non-compact groups.  This is because the gravitino is 
non-physical with no kinetic energy, so that we do not need 
the definite signature for the kinetic term.}    

The covariant derivative on the gravitino has also a 
minimal coupling to the $~B\-$field: 
$$ \li{ & \calR\du{\m\n r_1\cdots r_n} A  
   \equiv D_\m(\o, B) \psi\du{\n r_1\cdots r_n} A 
     - D_\n(\o, B)  \psi\du{\m r_1\cdots r_n} A \cr  
& ~~~~~  \equiv \big[ + \partial_\m \psi\du{\n r_1\cdots r_n} A
       + \frac 14 \o\du\m{t u} \g_{t u} 
     \psi\du{\n r_1\cdots r_n} A + n\o\du{\m (r_1 |} s 
     \psi\du{s | r_2\cdots r_n)} A   \cr 
& ~~~~~  ~~~~~ + g B\du\m I (T^I)^{A B} \psi\du{\n r_1\cdots r_n} B\, \big] 
      - {\scst ( \m \leftrightarrow \n)} ~~, 
&(2.5) \cr } $$ 
in addition to the non-trivial Lorentz connection term 
as in \ad.  The field strengths $~F, ~ 
G$~ and $~H$~ are defined by 
$$ \li{ & F\du{\m\n} I \equiv \partial_\m A\du\n I 
     - \partial_\n A\du\m I + g f^{I J K} A\du\m J A\du \n K ~~, \cr 
& G\du{\m\n} I \equiv \partial_\m B\du\n I 
     - \partial_\n B\du\m I + g f^{I J K} B\du\m J B\du \n K ~~, \cr   
& H\du{\m\n} I 
     \equiv  \partial_\m C\du\n I -  \partial_\n C\du\m I 
         + 2 g f^{I J K} B\du{\[ \m} J C\du{\n\]} K 
      \equiv D_\m C\du\n I - D_\n C\du\m I ~~.   
&(2.6) \cr} $$ 
     
We adopt the 1.5-order formalism, so that the Lorentz connection 
$~\o$~ in the lagrangian is regarded as an independent variable,
yielding its field equation\footnotes4{We use  the 
symbol $~{\scst\eqdot}$~ for a field equation distinguished from 
an algebraic identity.} 
$$ \li{ & \o_{m r s} \eqdot \Hat \o_{m r s}
     \equiv \frac  12 (\Hat C_{m r s} - \Hat C_{m s r} + \Hat C_{s r m} ) ~~,\cr 
& \Hat C\du{\m\n} m \equiv 
     \partial_\m e\du\n m - \partial_\n e\du\m m 
     - ( 2n+1) (\Bar\psi\du{\m(n)} A\g^m\psi\du\n{(n) A}) ~~. 
&(2.7) \cr } $$ 
Or equivalently, the torsion tensor is   
$$ \li{ & T\du{\m\n} m \eqdot + (2n+1) (\Bar\psi\du{\m(n)} A \g^m
     \psi\du\n{(n) A} ) ~~,  
&(2.8) \cr } $$ 
with the dependence on $~n$, in agreement with \ad.   

Our action $~I_0$~ is invariant under hypersymmetry 
$$ \li{ & \d_Q e\du\m m = + (2n+1) \big(\Bar\e\du{(n)} A \g^m 
     \psi^{(n) A}\big) ~~, \cr 
& \d_Q \psi\du{\m m_1\cdots m_n} A 
      = + \partial _\m\e\du{m_1\cdots m_n} A 
    + \frac 14 \Hat\o\du\m{r s} \g_{r s} \e\du{m_1\cdots m_n} A
     + n \Hat\o\du{\m(m_1 | } t\e\du{t|m_2\cdots m_n)} A \cr 
& {\hskip 1.0in} + g (T^I)^{A B} B\du\m I \e\du{m_1\cdots m_n} B   
     - g(T^I)^{A B} \g^\n \e \du{m_1\cdots m_n} B \Hat H\du{\m\n} I \cr  
& {\hskip 0.8in} \equiv + D_\m (\Hat\o, B) \e\du{m_1\cdots m_n}A 
     - g(T^I)^{A B} \g^\n \e \du{m_1\cdots m_n} B \Hat H\du{\m\n} I 
     ~~, \cr 
& \d_Q A\du\m I = - (T^I)^{A B} \big(\Bar\e\du{(n)}A \g_\m 
        \l^{(n) B} \big) ~~, \cr 
& \d_	Q B\du\m I 
      = - (T^I)^{A B} \big( \Bar\e\du{(n)} A \g^\n 
     \calR\du{\m\n}{(n)B} \big)   
     - h (T^I)^{A B} \big(\Bar\e\du{(n)} A \g_\m \l^{(n)B} \big) ~~, \cr 
& \d_Q C\du\m I = + (T^I)^{A B} \big(\Bar\e\du{(n)} A 
       \psi\du\m{(n) B} \big) 
       - h(T^I)^{A B} \big(\Bar\e\du{(n)} A \g_\m \l^{(n) B} \big) ~~, \cr 
& \d_Q \l\du{(n)} A = + \frac 12 (T^I)^{A B}  
    \g^{\m\n} \e\du{(n)} B 
     \big(2 F\du{\m\n} I + G\du{\m\n} I + \Hat H\du{\m\n} I\big) 
     - \frac 12(2n+1) \big(\Bar\e\du{(n)'}B\g^\m\psi\du\m{(n)'B} \big)
      \l\du{(n)} A \cr
&  {\hskip 0.65in} + 2 (T^I)^{A B} (T^I)^{C D} \big(\g_{\[\m} 
      \psi\du{\n\] (n)} B \big) 
    \big(\Bar\e\du{(n)'} C \g^\m\Tilde\calR^{\n (n)' D} \big) ~~, 
&(2.9) \cr } $$ 
where $~\Hat H\du{\m\n} I $~ is the `hypercovariantization' 
of $~H\du{\m\n} I$:
$$\li{ \Hat H\du{\m\n} I 
     \equiv \, & H\du{\m\n} I 
     - \Tab (\Bar\psi\du{\m(n)} A \psi\du\n{(n) B}) 
     + 2 h \Tab (\Bar\psi\du{\[\m| (n)} A \g_{| \n\]} \l^{(n)B}) 
    {~~, ~~~~~}  
&(2.10) \cr } $$ 
similar to supercovariantization 
\ref\pvn{{\it See, e.g.,} P.~van Nieuwenhuizen, \prep{68}{81}{189}.}%
\ref\ggrs{S.J.~Gates, Jr., M.T.~Grisaru, M.~Ro\v cek 
and W.~Siegel, {\it `Superspace'}  (Benjamin/Cummings,
Reading, MA 1983).}%
\ref\wb{J.~Wess and J.~Bagger, {\it `Superspace and Supergravity'}  
(Princeton University Press, 1992).}.   
The $~\Tilde\calR$~ is the Hodge dual of the 
gravitino field strength (2.5):  
$$ \li{ & \Tilde\calR\du{\m(n)} A 
     \equiv + \frac 12 e^{-1} \e\du\m{\r\s} \calR\du{\r\s(n)} A ~~. 
&(2.11) \cr  } $$ 
In accordance with (2.1), the parameter of hypersymmetry 
$~\e$~ should have an extra constraint:
$$ \li{ & \g^{m_1} \e\du{m_1 m_2\cdots m_n} A = 0 ~~. 
&(2.12) \cr } $$ 

The confirmation of action invariance under hypersymmetry is 
very similar to usual supergravity \pvn\ggrs\wb\  
or the case for $~n=0$~ in \nralephnull.  There are in total 
ten different categories of $~g$~ or $~h\-$dependent 
terms (sectors):  (i) $~g\psi G$,~
(ii)  $~ g \calR H$, ~ (iii)  $~g \psi^e \calR$, ~ (iv)  $~g h F \l $, ~
(v) $~ g h G \l$, ~ (vi) $~g h H \l $,~(vii) $~g h \psi \l^2$, ~
(viii)  $~ g h \psi^2 \l$~,  (ix)  $~h \psi^2 \calR $, ~(x) $~g \psi \calR \l$.   
Among these, the antisymmetry of $~T^I$~ is frequently 
used in the sectors (ii), (iii), (iv), (v), (vi), (viii) and (ix).   Other 
cancellation patters are more or less parallel to \nralephnull.  
    
The closure of gauge algebra is similar to \nralephnull, as  
$$ \li{ & \[ \d_Q (\e_1) , \d_Q(\e_2) \] 
     = \d_P(\xi) + \d_{\rm G}(\xi) + \d_Q(\e_3) + \d_{\rm L}(\l^{r s}) 
       + \d_\L + \d_{\Tilde \L} + \d_{\doubletilde\L} ~~, \cr 
& \xi^m \equiv + (2n+1) (\Bar\e\du{2(n)} A \g^m \e\du1{(n) A} ) ~~, 
 ~~~~~ \e\du{3(n)}A \equiv - \xi^\m\psi\du{\m(n)} A ~~, \cr 
& \l^{r s} \equiv + \xi^\m\Hat\o\du\m{r s} 
      - 2 (2n+1) g \Tab (\Bar\e\du{1(n)}A \e\du 2{(n) B} ) 
     \Hat H^{r s I} ~~, \cr 
& \L^I  \equiv - \xi^\m A\du\m I ~~, ~~~~
      \Tilde\L^I  \equiv - \xi^\m B\du\m I ~~, ~~~~
     \doubletilde\L^I \equiv - \xi^\m C\du \m I 
     - \Tab (\Bar\e\du{1(n)} A \e\du2{(n) B}) 
      {~~, ~~~~~ ~~~~~}     
&(2.13) \cr } $$ 
where $~\d_P, ~\d_{\rm G}, ~\d_Q, ~\d_{\rm L}$~ are 
the translation, general coordinate, hypersymmetry and 
local Lorentz transformations, while 
$~\d_\L, ~\d_{\Tilde \L}$~ and 
$~\d_{\doubletilde \L}$~ are the $~G\-$gauge transformations
$$ \li{ & \d_\L A\du\m I \equiv \partial_\m \L^I 
    + g f^{I J K} A\du\m J \L^K~~, \cr 
& \d_{\Tilde\L} B\du\m I \equiv \partial_\m \Tilde\L^I 
    + g f^{I J K} B\du\m J \Tilde\L^K~~, \cr 
& \d_{\doubletilde\L} C\du\m I \equiv \partial_\m \doubletilde\L^I 
    + g f^{I J K} B\du\m J \doubletilde \L^K~~, 
&(2.14) \cr } $$ 
As usual, the closure of on-shell hypersymmetry can be 
confirmed by the use of the field equations: 
$$ \li{ & F\du{\m\n} I \eqdot 0 ~~, ~~~~
       G\du{\m\n} I \eqdot 0 ~~, ~~~~
     \Hat H\du{\m\n} I \eqdot 0 ~~, 
&(2.15\rma)  \cr 
& \calR\du{\m\n} A \eqdot 0 ~~. 
&(2.15\rmb)  \cr } $$

The special case when the gravitino carries no 
group index recovers the result by Aragone-Deser \ad.  
Also, the case $~n=0$~ with the adjoint group index maintained is the 
$~\alephnull\-$supergravity with the spin 3/2 gravitino 
with the adjoint index $~{\scst A~=~I}$~ \nralephnull.  

There is a technical subtlety associated with the field equation 
of  $~\o$~ and the torsion (2.8), which was not stressed well in \ad.    
The $~\o\-$field equation from our lagrangian is 
\vbox{
$$ \li{ \fracmm{\d I_0} {\d\o\du\m{r s} } 
      = \, & - \frac14 e T\du{r s}\m 
              -  \frac 12   e e\du{\[ r } \m T_{s\]} 
      - \frac 18 \e^{\m\r\s}  (\Bar\psi\du{\r (n)}A \g_{r s}  
      \psi\du\s {(n) A} ) \cr 
& - \frac 12 n \e^{\m\r\s} (\Bar\psi\du{\r r(n-1)} A \psi\du{\s s}{(n-1) A}) 
      \eqdot 0 ~~,  
&(2.16) \cr } $$ 
}
where $~T_\m \equiv T\du{\m \n} \n$.  By multiplying this by 
$~e\du s \m$, we get the expression for $~T_\m$~ which in turn is 
substituted into (2.16) to eliminate the $~T_\m\-$term, as 
$$\li{ & T\du{\r \s} m = + (\Bar\psi\du{\r (n)} A \g_m \psi\du\s {(n) A} \big) 
      - 6 n e^{-1} \e^{\t\l\o} e\du{\[ \r|} m \big(\Bar\psi\du{\t | \s| }{(n-1) A}  
     \psi\du{\l | \o\] }{(n-1) A} \big)  {~~. ~~~~~ ~~ } 
&(2.17) \cr } $$   
Interestingly enough, the last term antisymmetric 
in $~{\scst \[\r \s \o\]}$~ can be shown to be proportional to 
the first term in (2.17) by the peculiar algebra 
$$ \li{ &  3 e^{-1} \e^{\t\l\o} 
     e\du{\[ \r|} m \big(\Bar\psi\du{\t | \s|(n-1)} A 
     \psi\du{\l | \o\] }{(n-1) A} \big) 
     = - \frac 12 e^{-3} \e^{\t\l\o} \e_{\r\s\o} \e^{\m\n\psi} 
          e\du\m m \big(\Bar\psi\du{\t\n(n-1)} A 
         \psi\du{\l\psi}{(n-1) A} \big) \cr 
& = + e^{-1} \e^{m \t \l} \big(\Bar\psi\du{\r\t(n-1)} A 
      \psi\du{\s\l}{(n-1) A} \big) 
    =  + \big(\Bar\psi\du{\r\t(n-1)} A \g^{m \t\l} 
        \psi\du{\s\l}{(n-1) A} \big) \cr
& = + \big(\Bar\psi\du{\r\t(n-1)} A \g^{m \t} 
          \g^\l \psi\du{\s\l}{(n-1) A} \big) 
     - \big(\Bar\psi\du{\r\t(n-1)} A\g^m\psi\du\s{\t(n-1) A}\big) 
    + \big(\Bar\psi\du{\r\t(n-1)} A \g^\t \psi\du\s{m(n-1) A}\big) \cr 
&  = - \big(\Bar\psi\du{\r(n)} A\g^m \psi\du{\s}{(n) A} \big) ~~. 
&(2.18) \cr } $$ 
Here use is made of the important constraint (2.1) for dropping the first
and last terms in the penultimate line in (2.18).  This combines 
the last term in (2.17) with its first term, with its coefficient changed from
unity to $~(2n+1)$.          

It is sometimes convenient to note the difference in the 
transformation of the Lorentz connection between 
the first and second-order formalisms:  
\vskip -0.41in 
$$\li{ \hbox{(i)}~  &\hbox{First-Order:}  ~~ \d_Q \o_{\m r s} = 0 ~~, 
&(2.19\rma) \cr 
\hbox{(ii)}~ &\hbox{Second-Order:}   ~~ \d_Q \Hat\o_{\m r s} 
      = \frac 12 (n+1) \big[\, 2(\Bar\e^{(n) A} 
      \g_{\[r} \calR\du{s\] \m (n)} A ) 
    - (\Bar\e^{(n) A} \g_\m \calR\du{r s (n)} A ) \big] 
      {~. ~~~~~ ~~~~~ ~~} 
&(2.19\rmb) \cr } $$ 
\vskip -0.16in 

\noindent  
These two are equivalent to each other on-shell under (2.15b),   
as in the 4D case \pvn\ggrs\wb.

We mention the consistency of gravitino field equation
\ref\dz{S.~Deser and B.~Zumino, \pl{62}{76}{335}.} 
in  our system, associated with so-called `Velo-Zwanziger disease'
\vz.  This is about the problem with the divergence of the gravitino 
field equation on curved backgrounds or non-trivial  field strengths
frequently encountered in 4D or higher.   Fortunately in 3D, the
covariant divergence of gravitino field equation (2.15b) vanishes, as
desired:    
$$ \li{ 0 \eqques D_\m \! & 
     \big( \frac 12 \e^{\m\r\s}
     \calR\du{\r\s m_1\cdots m_n} A \big)  
      = + \frac 18 \e^{\m\n\r} (\g_{r s} \psi\du{\r m_1\cdots m_n} A ) 
      R\du{\m\n}{r s} (\o) \cr 
& + \frac 12 n \e^{\m\n\r} \psi\du{\r t (m_2\cdots m_n|} A 
      R\du{\m\n |m_1)} t (\o) 
    + \frac 12 g \e^{\m\n\r} (T^I)^{A B} \psi\du{\r m_1\cdots m_n} B 
     G\du{\m\n} I \eqdot 0 {~~. ~~~~~ ~~~~~}    
&(2.20) \cr } $$ 
In particular, $~G\du{\m\n} I \eqdot 0~$ (2.15a) and the 
dreibein field equation 
$$\li{ & R _{\m\n} (\o) \eqdot 0 ~~ \Longleftrightarrow
      ~~ R\du{\m\n}{r s} (\o) \eqdot 0 
&(2.21) \cr } $$ 
are used.  The latter is associated with  
the peculiar identity in 3D: 
$$\li{& R\du{\m\n}{\r\s}(\o) 
      \equiv + 4 \d\du{\[\m }{\[\r } R\du{\n\] }{\s\] } (\o) 
      - \d\du{\[\m}{\[\r } \d \du{\n\] }{\s\] } R(\o)  ~~,  
&(2.22) \cr } $$ 
due to the vanishing of the conformally invariant Weyl tensor 
$~C\du{\m\n\r}\s\equiv 0$~ \djt.  

As usual in 3D, the coefficient of the Chern-Simons  term should be 
quantized \djt\ for a compact gauge group whose $~\pi_3\-$homotopy
mapping is  non-trivial, {\it e.g.,} 
$$ \li{& \pi_3(G) 
= \cases{  \ZZ  & (\hbox{for}  ~$G = A_n, ~B_n,  ~
     C_n, ~D_n ~~(n\ge 2, ~G\neq D_2), 
      ~G_2, ~F_4, ~E_6, ~E_7, ~ E_8) { ~,~~} $   \cr  
\ZZ \oplus \ZZ & (\hbox{for} $~G = SO(4))$~, \cr 
0 & (\hbox{for} ~$G = U(1))~. $ \cr } \cr 
\noalign{\vskip -0.5in} \cr & ~~~ 
&(2.23) \cr } $$  
Especially for a compact gauge group with $~\pi_3(G) = \ZZ$, 
the quantization condition is \djt\footnotes5{The wrong power of 
$~{\scst g}$~ in (2.12) of \nralephnull\ is corrected in (2.24).}   
$$ \li{ & g k = 8\pi h ~~~~~ (k= 0, ~\pm 1, ~\pm 2,  ~\cdots) ~~. 
&(2.24) \cr } $$ 
Under this condition, the two initially independent constant 
$~g$~ and $~h$~ are now proportional to each other.

\bigskip\bigskip\medskip 


\leftline{\bf 3.~~Additional Chern-Simons Terms}   

Due to the peculiar property of 3D, we can further add some 
Chern-Simons terms to our initial lagrangian $~\Lag_0$.  
The first example is the Lorentz Chern-Simons term  
$~I_{R \o} \equiv \int d^3 x \, \Lag_{R \o}$~  \djt, where 
$$ \li{ \Lag_{R \o} \equiv \, & 
     + \frac 12 \a \e^{\m\n\r} \big[\,  R\du{\m\n}{r s} (\o) \o_{\r r s} 
      + \frac 13 \o\du{\m r} s \o\du{\n s} t \o\du{\r t} r \, \big]  ~~. 
&(3.1) \cr } $$ 
Here $~\a$~ is a real constant.  This lagrangian was also presented 
in the context of topological massive gravity \djt.    

The $~\o\-$field equation is now modified by $~\Lag_{R \o}$~ 
with $~\a$, as 
$$ \li{ \fracmm{\d I_{\rm tot}} {\d\o\du\m{r s} } 
      = \, & - \frac14 e T\du{r s}\m 
              -  \frac 12   e e\du{\[ r } \m T_{s\]} 
      - \frac 18 \e^{\m\r\s}  (\Bar\psi\du{\r (n)}A \g_{r s}  
      \psi\du\s {(n) A} ) \cr 
& - \frac 12 n \e^{\m\r\s} 
      (\Bar\psi\du{\r r(n-1)} A \psi\du{\s s}{(n-1) A}) 
      + \a \e^{\m\r\s} R_{\r\s r s} (\o) \eqdot 0 ~~.  
&(3.2) \cr } $$ 
However, the new term is proportional to the 
Riemann tensor $~R\du{\m\n}{r s}(\o)$, and it vanishes because of (2.21) 
and the dreibein field equation (2.21) which is not modified.
In other words, the last $~\a\-$dependent term in (3.2) 
vanishes, leaving the original algebraic field equation for $~\o$.  
Accordingly, the torsion (2.8) stays intact, and we can 
still use the first or 1.5-order formalism for $~\o$.  Consequently, the 
invariance of the total action $~I_{\rm tot} \equiv I_0 + I_{R \o}$~ 
is still valid despite the presence of $~I_{R \o}$.   

This is based on the 1.5-order formalism, but  the invariance $~\d_Q
I_{\rm tot}=0$~ is  much more transparent in the first-order formalism. 
This is because in the first-order formalism, the Lorentz connection is
invariant under hypersymmetry as in (2.19a), and therefore the 
invariance $~\d_Q I_{R \o} =0$~ is manifest.

Note that the fact $~R\du{\m\n} {r s} (\o) \eqdot 0$~ does {\it not} 
make the Lorentz-Chern-Simons term (3.1) trivial, because of its 
last term $~\o \wedge \o\wedge \o$, as in the usual Chern-Simons 
term in 3D \djt.  Additionally, since the Lorentz group $~SO(2,1)$~ in 
3D is noncompact, there  is no quantization for the coefficient 
$~\a$~ \djt.  

Another interesting Chern-Simons term is of the $~G B\-$type  
$~I_{G B} \equiv \int d^3 x\, \Lag_{G B}$, where  
$$\li{ & \Lag_{G B} 
   \equiv \frac 12 \b \e^{\m\n\r} \big(
     G\du{\m\n} I B\du \r I - \frac 1  3 g f^{I J K} 
    B\du\m I B\du \n J B\du \r K \big) ~~. 
&(3.3) \cr } $$ 
In order to maintain the invariance of the total action
$~ I_{\rm tot} \equiv I_0 + I_{R \o} +I_{G B}$~  under 
hypersymmetry, we need to shift the $~C\-$field 
transformation rule by $~\d_Q B$, as 
$$ \li{ \d_Q C\du \m I 
     = \, & + (T^I)^{A B} \big(\Bar\e\du{(n)} A 
     \psi\du\m{(n) B} \big) - h (T^I)^{A B} 
     \big(\Bar\e\du{(n)} A \g_\m \l^{(n) B} \big) 
   - g^{-1} \b  ( \d_Q B\du\m I ) {~~.  ~~~~~ ~~}
&(3.4) \cr } $$ 
Since the $~C\-$field appears nowhere in $~\Lag_{\rm tot}$~ other than 
the $~C G\-$term, and the variation of $~I_{G B}$~ has only 
$~\d_Q B$, the modification above is sufficient for the invariance of 
$~I_{\rm tot}$.  The $~B\-$field equation seems to get modified 
by the $~\b\-$term, but actually not, because of the $~C\-$field 
equation $~G\du{\m\n}I \eqdot 0$.    

The quantization of the $~\b\-$coefficient for any gauge group 
with $~\pi_3(G) = \ZZ$~ is 
$$ \li{ & g^2 \ell = 8 \pi \b ~~~~~( \ell = 0 , ~\pm 1, ~\pm 2, ~\cdots) ~~. 
&(3.5) \cr } $$ 
This restricts the originally independent constant $~\b$~ 
to be proportional to $~g$.

\bigskip\bigskip\medskip 


\leftline{\bf 4.~~Concluding Remarks}   

In this paper, we have presented $~^\forall N\-$extended 
hypergravity with the gravitino of spin $~n+3/2$~ in an arbitrary real
representation satisfying the condition $~ (T^I)^{A B} = - (T^I)^{B A}$~ of
any gauge group $~G$.  Since such representation can be 
limitlessly large, we have arbitrarily large extended  local
hypersymmetries, which we call `$\alephnull\-$hypergravity'.   The
number $~N$~ of hypersymmetries is specified by the dimensionality 
of the representation of the index $~{\scst A}$.  For example, 
the adjoint representation $~{\scst A~=~I}$~ 
yields $~N = \hbox{dim}\, G$, or the vectorial representation 
$~{\scst A~=~a}$~ of $~G= SO(M)$~ yields $~N = M$.  

Our result is a generalization of the work by Aragone-Deser for
gravitino with arbitrarily large half-integral spins \ad\ to the system 
where  the gravitino is in an arbitrary real representation such that the
generators are antisymmetric.   Compared with \ad, our system  has also
additional structures, such as the topological $~C G\-$term and the 
hypersymmetric Chern-Simons term.  When the gauge group is trivial as
$~G =I$, our graviton-gravitino sector is reduced to \ad, while the $~C
G$~ and the Chern-Simons term still remains as an Abelian case.  On the
other hand, if we keep $~G$~ to be non-trivial, while putting $~n =0$~  
with the spin 3/2 gravitino, we get the generalization of
$~\alephnull\-$supergravity \nralephnull\ for more general 
representations than the adjoint representation.  

We have further generalized our system, by adding  two more
Chern-Simons term $~I_{R \o} $~ and $~I_{G B}$.  We have seen that the
hypersymmetric invariance  of the total  action $~I_{\rm tot} \equiv I_0 +
I_{R\o} + I_{G B}$~ is  restored, by a slight modification of the $~C\-$field
transformation  rule.  Even though the field equations imply that 
$~R\du{\m\n} {m n}(\o)\eqdot 0$~ and $~G\du{\m\n}I \eqdot 0$~
on-shell, the $~\o\wedge\o\wedge\o$~ for the former or $~B\wedge
B\wedge B$~  for the latter has non-trivial contribution as the surface
term,  as usual in Chern-Simons theory.   We have also seen interesting 
relationships among the coupling constants $~g$~ and $~h$~ by the 
quantization of the coefficients of the relevant Chern-Simons terms
for any gauge group with $~\pi_3(G) = \ZZ$.  

In this paper, we have given Chern-Simons formulations with no  more
fundamental bases.  However, it may well be the case that our theory  
has foundations, such as superparticle,  superstring or 
supermembrane theory.  As a matter of fact, it has been known for some
time that a Chern-Simons theory in 10D can be derived from the second
quantization of superparticle theory 
\ref\kallosh{R.~Kallosh, 
{\it `$D = 10$ Supersymmetric Chern-Simons Gauge Theory'}, 
Talk given at {\it `Strings and Symmetries'}, 
in {\it `Strings:  Stony Brook 1991'}, pp.~341-354, hep-th/9111010.}.  
Therefore, it is not too far-fetched to expect that our 
$~\alephnull\-$hypergravity has also such extended objects 
as its foundation.  

Even though the field strengths in our system vanish 
upon their field equations, they might have more significance 
than expected.  In fact, a similar situation is 
found in so-called loop quantum gravity theory
\ref\lqg{A.~Ashtekar, V.~Husain, C.~Rovelli, J.~Samuel and L.~Smolin,
\cqg{6}{89}{L185};   
L.~Smolin, {\it `Loop Representation for Quantum Gravity in 2+1
Dimensions'}, in {\it `John's Hopkins Conference on Knots, 
Tolopoly and Quantum Field Theory'} (World Scientific, 1989); 
C.~Rovelli and L.~Smolin, \np{331}{90}{80};  
L.~Smolin, {\it `An Invitation to Loop Quantum Gravity'},  
\hepth{0408048}.},    
or more generally, in topological field theories \tft,  where vanishing
field strengths play non-trivial roles due to topological effects on the 
boundary.  In our system, it is very peculiar that the local gauge
symmetry can be arbitrarily large, and accordingly local
hypersymmetry can be also  arbitrarily large  ($\alephnull$) as well.

\bigskip\bigskip\medskip 


We are grateful to S.~Deser for important discussions.  
This work is supported in part by NSF Grant \# 0308246.

\bigskip\bigskip\bigskip

\vfill\eject 

\immediate\closeout\rfile\writestoppt
\baselineskip=14pt\centerline{{\bf References}}%
\bigskip{\frenchspacing%
\parindent=20pt\escapechar=` \input refs.tmp\vfill\eject}%
\nonfrenchspacing

\vfill\eject

\end{document}